\documentclass[]{spie}  

\usepackage{amsmath,amsfonts,amssymb}
\usepackage{graphicx}
\usepackage[colorlinks=true, allcolors=blue]{hyperref}

\title{Astrometric and Wavelength Calibration of the NIRSpec Instrument during Commissioning using a model-based approach}

\author[a]{Nora L\"utzgendorf}
\author[b]{Giovanna Giardino}
\author[c]{Catarina Alves de Oliveira}
\author[d]{Peter Zeidler}
\author[c]{Pierre Ferruit}
\author[e]{Peter Jakobsen}
\author[d]{Nimisha Kumari}
\author[a]{Timothy Rawle}
\author[a]{Stephan M. Birkmann}
\author[a]{Torsten B\"oker}
\author[f]{Charles Proffitt}
\author[a]{Marco Sirianni}
\author[a]{Maurice Te Plate}
\author[f]{Sangmo Tony Sohn}

\affil[a]{European Space Agency, c/o STScI, 3700 San Martin Drive, Baltimore, MD 21218, USA}
\affil[b]{ATG Europe for the European Space Agency, Noordwijk, The Netherlands}
\affil[c]{European Space Agency, ESAC, Madrid, Spain}
\affil[d]{AURA for the European Space Agency, STScI, Baltimore, USA}
\affil[e]{Cosmic Dawn Center, Niels Bohr Institute, University of Copenhagen, Denmark}
\affil[f]{Space Telescope Science Institute, Baltimore, USA}

\authorinfo{Further author information: E-mail: nora.luetzgendorf@esa.int}

\newcommand{\mum}{\,{\mu {\rm m}}}

\begin{document} 
\maketitle

\begin{abstract}
The NIRSpec instrument for the James Webb Space Telescope (JWST) is a highly versatile near-infrared spectrograph that can be operated in various observing modes, slit apertures, and spectral resolutions. Obtaining dedicated calibration data for all possible combinations of aperture and disperser is an intractable task. We have therefore developed a procedure to derive a highly realistic model of the instrument's optical geometry across the entire field of view, using calibration data acquired through only a subset of NIRSpec apertures, which nevertheless allows the light paths within the spectrograph to be accurately computed for all apertures and all observing modes. This parametric instrument model thus provides the basis for the extraction of wavelength-calibrated spectra from any NIRSpec exposure, regardless of observing mode or aperture used. Optimizing the NIRSpec instrument model and deriving its final wavelength and astrometric calibration was one of the most crucial elements of the NIRSpec commissioning phase. Here, we describe the process of re-fitting the NIRSpec instrument model with in-orbit commissioning data, and present its final performance in terms of wavelength accuracy and astrometric calibration. 
\end{abstract}


\section{INTRODUCTION}
\label{sec:intro}  

The Near Infrared Spectrograph (NIRSpec) onboard the James Webb Space Telescope (JWST) has four main scientific observing modes: Fixed Slits (FS), Multi-object spectroscopy (MOS), Integral Field Unit (IFU) mode and Bright Object Time Series (BOTS). Operating in the near infrared in the wavelength range of $0.6 - 5.3\mum$, and using a collection of gratings and a prism that allow to take spectra in low (R=100), medium (R=1000) and high (R=2700) resolution, NIRSpec is designed to take up to 200 spectra at once, for example to study high-redshift galaxies and the early universe. A detailed description of the NIRSpec design, observing modes, and scientific use cases can be found in [\citenum{Jakobsen22,Ferruit22,Boeker22,Birkmann22}].

Because of its complexity especially in the MOS mode (which has $\sim$ 250000 individual shutters), calibrating the wavelength and astrometric solutions using a static method for each individual shutter and disperser is impossible. In addition, the grating wheel assembly (GWA), which is the center piece of the internal light path, has limited angular positioning repeatability. This means that whenever the GWA returns to the same grating, it does so with a slightly different position, causing the spectra on the detector to shift by small amounts, both in the dispersion and cross-dispersion direction.

\begin{figure} [ht]
	\begin{center}
		\begin{tabular}{c} 
			\includegraphics[width=\textwidth]{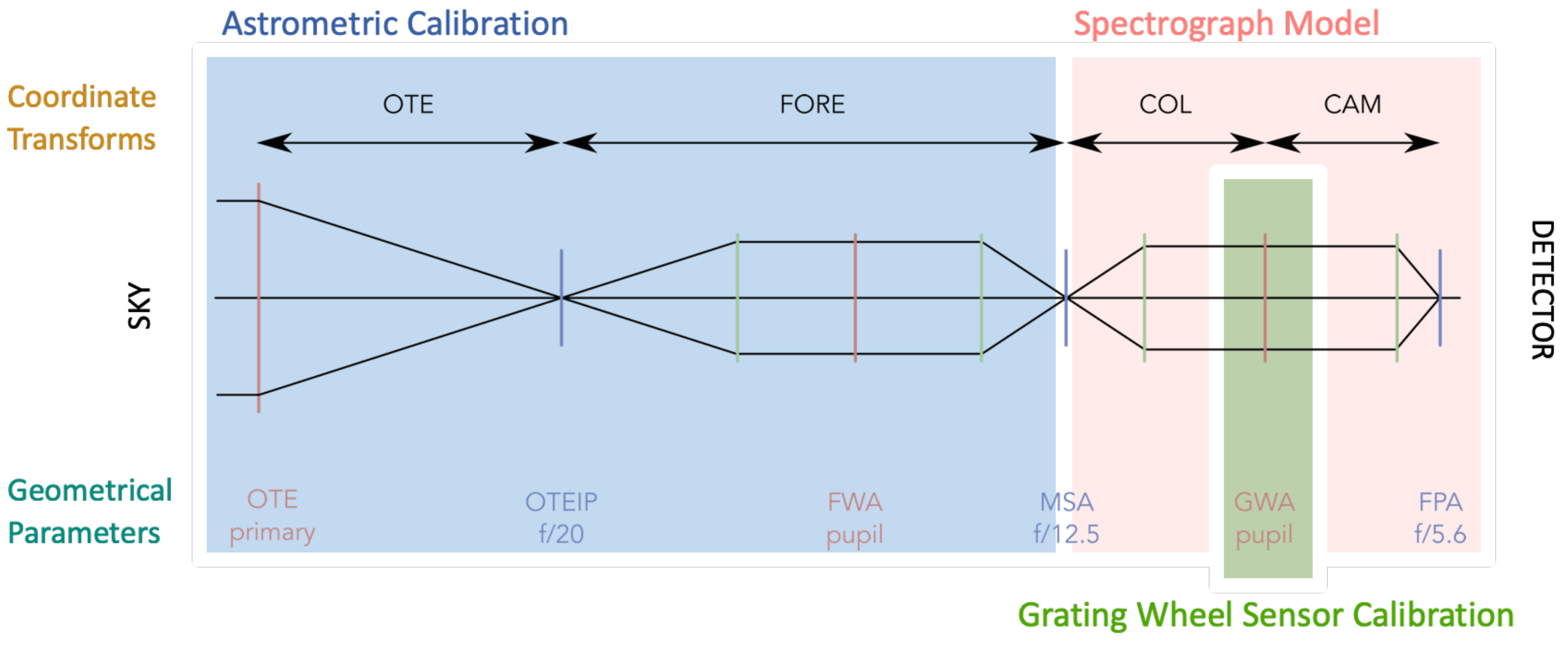}
		\end{tabular}
	\end{center}
	\caption{Schematic overview of the NIRSpec instrument model. The purpose of the model is to describe the light path through the instrument from sky to the detector and vice versa. It is composed of coordinate transforms and geometrical parameters and splits into three main parts: The spectrograph model describing the path between the MSA and the detector, the grating wheel sensor calibration and the astrometric calibration.} 
	\label{fig:scheme} 
\end{figure}

\begin{figure} [ht]
	\begin{center}
		\begin{tabular}{c} 
			\includegraphics[width=\textwidth]{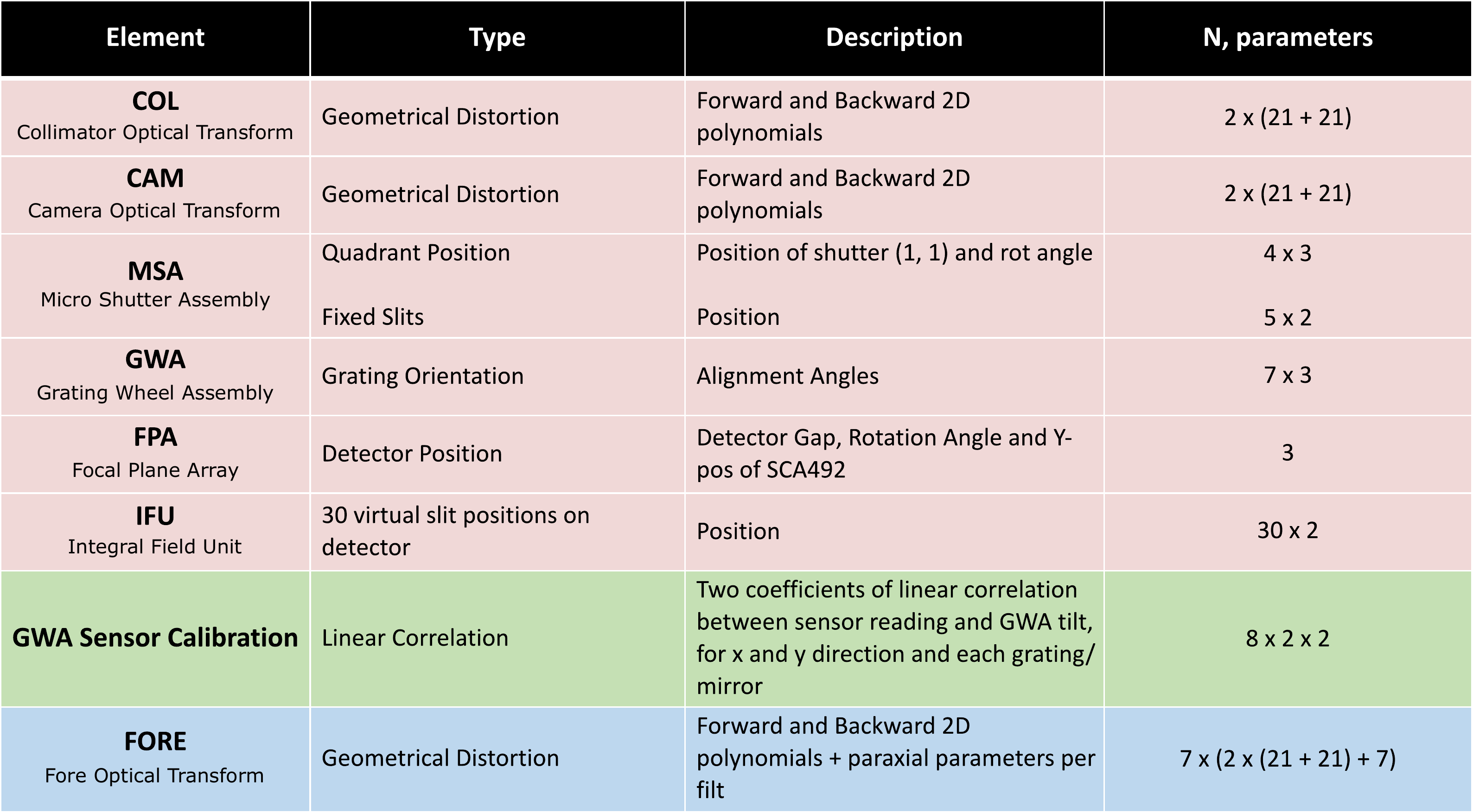}
		\end{tabular}
	\end{center}
	\caption{Overview of all model parameters that are fit in commissioning. The different parts of the model are highlighted in the corresponding colors as in figure \ref{fig:scheme}. The total number of parameters fit in commissioning is $>$ 900.} 
	\label{fig:pars} 
\end{figure}

Therefore, we have developed a procedure that uses calibration data acquired for a limited subset of the NIRSpec apertures to derive a highly realistic model of the instrument's optical geometry. The NIRSpec instrument model consists of a collection of coordinate transforms between the various optical planes illustrated in Fig.\,\ref{fig:scheme} (OTE, FORE, COL, CAM, IFU-FORE, IFU-POST - see Fig.\,\ref{fig:pars} for acronyms), each represented by a paraxial transformation (rotation, magnification and offset) and a 5th-degree 2-D polynomial representing the local distortions. The model also relies on a set of geometrical parameters that capture the physical properties of key optical elements (MSA, IFU slicer, GWA and FPA), for example the locations of the individual quadrants in the micro shutter assembly (MSA) plane, the precise orientation of the dispersers in the GWA, or the location and orientation of the detectors in the FPA.

The model is split up into three parts that require independent calibration, but build on each other. The first is the spectrograph or internal model. It describes the light path from the MSA plane to the detector and back. It can be fit for all dispersers by using the internal calibration lamps, and thus does not require on-sky data. The procedure of fitting this part of the model is described in section \ref{sec:spec}. The second part of the model is the grating wheel sensor calibration described in [\citenum{Alves22}] and summarized in section \ref{sec:gwa}. The remaining FORE optical transforms are fit in the astrometric calibration step by registering the positions of stars on the detector with an astrometric catalogue as described in \ref{sec:fore}. Note that the OTE transform as shown in Fig.\,\ref{fig:scheme} is fixed to a pre-launch model, as it is not possible to fit this transform individually in flight. Figure \ref{fig:pars} lists all of the free parameters of the model, color coded by its three components. In total, the NIRSpec instrument model consists of $> 900$ free parameters.

\section{DATA}
\label{sec:data}

The final version of the instrument model was fit using in-flight commissioning data obtained as part of the 6-month long NIRSpec commissioning campaign between December 25, 2021 and July 15, 2022 (see [\citenum{Boeker22b}] for a detailed description). In addition, data taken during various ground campaigns as well as simulations were used to obtain absolute wavelength calibration of the internal lamps, and to develop procedures to prepare for commissioning. In the next sections, those campaigns and data are briefly summarized. 

\subsection{Instrument-level test campaigns}

The first two instrument test campaigns took place at Industrieanlagen-Betriebsgesellschaft mbH (IABG) facilities in Ottobrunn, Germany, in 2011 and 2013. The instrument was enclosed in a vacuum chamber cooled by gaseous Helium, and illuminated by external calibration light sources that used a cryo-mechanism to switch between flat-field illumination and a grid of pinholes. Using this set-up, an Argon lamp spectrum was used to perform the first spectral calibration of the instrument, and to assign absolute wavelengths to the internal calibration lamps (see [\citenum{Dorner16}] for more details). Furthermore, the pinhole mask was placed at the OTEIP plane of the instrument and used to derive a first calibration of the NIRSpec FORE optical transforms (astrometric calibration). Analysis of the data showed that the measurements were consistent with the as-designed optical transforms that were therefore kept as the baseline FORE optical transforms.

\subsection{Higher-level test campaigns}

After the delivery to NASA in 2013, NIRSpec was installed on the Integrated Science Instrument Module (ISIM), together with the other JWST science instruments. The fully assembled ISIM underwent two cryogenic test campaigns in 2014 (CV2) and 2015 (CV3) at the Goddard Space Flight Center (GSFC). Between CV2 and CV3, the instrument needed to be refurbished by replacing the MSA and FPA subsystems with new parts. This made the CV3 test campaign a crucial step in the calibration of NIRSpec, since for the first time, it was in its final flight configuration. Therefore, the spectrograph model needed to be completely redone. 

The final cryogenic test campaign was done after ISIM was integrated with the Optical Telescope Element (OTE), which together form the Optical Telescope Element and Integrated Science Module (OTIS). In 2017, OTIS was shipped to Johnson Space Center (JSC) and underwent extended cryogenic testing over the course of three months. For the NIRSpec instrument model, a subset of data was taken to apply a reduced model fit, mostly monitoring smaller differences in the geometric parameters (such as shifts in the MSA quadrants). This was a good test to see how the model changed after NIRSpec underwent environmental testing (vibe and acoustic).

\subsection{Simulations}

In addition to data taken by cryogenic test campaigns, we also produced a set of simulations using the instrument performance simulator (IPS, see [\citenum{Piqueras08}, \citenum{Piqueras10}]]) for NIRSpec which was developed by the Centre de Recherche Astrophysique de Lyon (CRAL) and delivered to ESA. This was particularly important for data that we could not reproduce with ground testing campaigns such as the astrometric calibration on sky. As the simulator uses the instrument model to generate the exposures, those could not be used to obtain a preliminary model fit for the FORE optics, but they were used to develop the routines necessary to perform the fit once on-sky data became available. 

\subsection{Commissioning Data} \label{subsec:commdata}

The final dataset used to perform the model fit was taken during the NIRSpec commissioning campaign. All exposures were processed with the NIRSpec Commissioning Team ramps-to-slopes pipeline that applies the following corrections: bias subtraction, reference pixel subtraction, linearity correction, dark subtraction and finally count-rate estimation, including jump detection and cosmic-ray rejection [\citenum{Birkmann22b}].

The data set needed to update the spectrograph part of the model was taken in Commissioning Activity Request (CAR) NIRSpec-041 (Instrument Model Update, Proposal ID 1132). This CAR used internal lamp exposures in imaging, MOS, FS and IFS modes. For imaging, the MSA was configured to a (3x3) checkerboard and a customized pattern with crosses, used for a first manual adjustment without the full fit. In order to determine the location of spectra through individual shutters (a.k.a. `traces'), a combination of four different slitlet configurations for each grating/lamp combination was used, in order to cover the detector area as evenly as possible. This data set also included IFU exposures and their dedicated `leakage' exposures, i.e. exposures with the IFU closed which only contain traces from failed open shutters (which can then be subtracted). Each combination of grating and MSA configuration was observed with the LINE, FLAT, and REF lamps at the same GWA position (i.e. without any intermittent GWA movement). Before changing to the next grating, the GWA was `spun around' once, and a REF lamp exposure was taken at each position. These extra observations were used for the grating wheel sensor calibration (see [\citenum{Alves22}] for a full list of data).

The exposures used for the remaining FORE portion of the model (a.k.a. astrometric calibration) were taken in CAR NIRSpec-019 (Astrometric Calibration, PID 1120). They consist of undispersed images of the JWST astrometric reference field (located in the Large Magellanic Cloud [\citenum{Sahlmann17}, \citenum{Anderson21}] through all seven NIRSpec filters, acquired with the GWA in the MIRROR position, the MSA configured to ALLOPEN, and the IFU aperture unblocked. For each filter, two images were obtained, separated by a half-shutter small angle maneuver (SAM) in order to be able to correct for obscuration by the MSA bars that separate the individual shutters.

Lastly, internal lamp exposures using the TEST lamp, the GWA in MIRROR, and four MSA configurations with customized test patterns (CROSS5-C, (1x1) and (3x3) checkerboards) were also taken as part of PID 1120. The purpose of these was to acquire internal exposures at the exact same grating wheel positions as the on-sky exposures, to provide the most accurate means to derive the transformation between the sky and the MSA plane. This approach was mostly a safeguard against the possibility that the existing sensor calibration derived from the PID 1141 data was not yet accurate enough. 

\section{SPECTROGRAPH MODEL}
\label{sec:spec}

The spectrograph (or internal) portion of the NIRSpec instrument model describes the light-path from the MSA plane to the detector and back. It can be calibrated on the ground and in space by using the NIRSpec internal calibration lamps. A full description of the fitting process and the first results obtained during the IABG tests can be found in [\citenum{Dorner16}]. 

\begin{figure} [ht]
	\begin{center}
		\begin{tabular}{c} 
			\includegraphics[width=\textwidth]{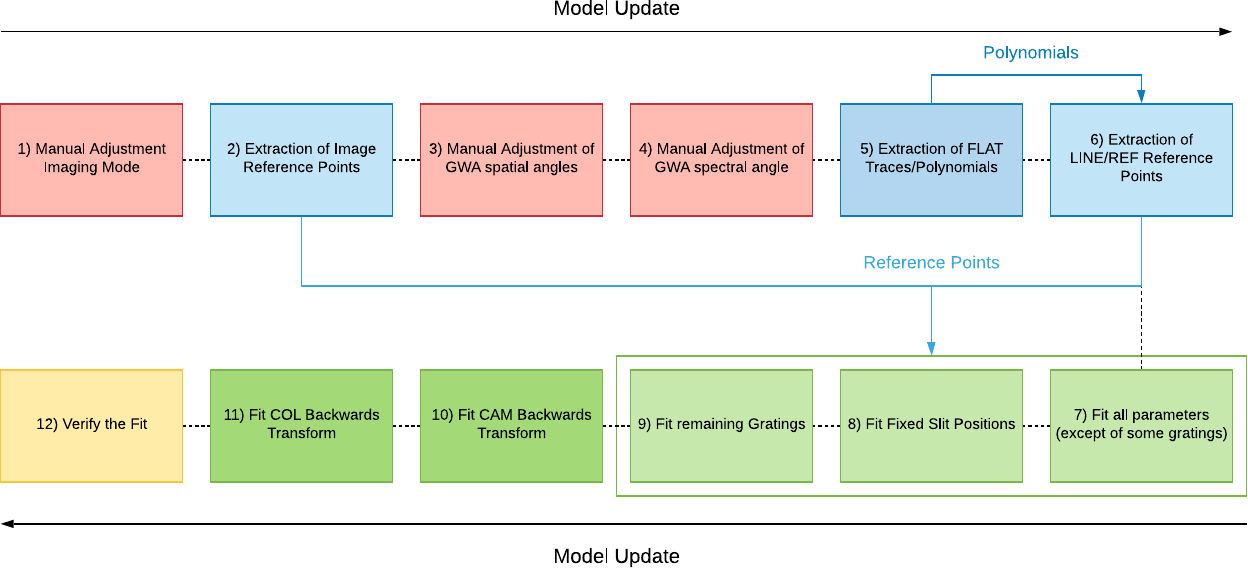}
		\end{tabular}
	\end{center}
\caption{Schematic workflow of the spectrograph model fit. Red boxes represent manual adjustments, blue boxes reference point creation and extractions, green boxes fits and the yellow box verification.} 
\label{fig:specfit} 
\end{figure}

The full fit procedure of the spectroscopic model is outlined in figure \ref{fig:specfit}. The red and blue boxes (manual adjustments and extraction of reference points) are described in sections \ref{subsec:man} and \ref{subsec:ext}, respectively. Green boxes represent the different steps when fitting to the extracted reference data, and the yellow box denotes the final verification step (both discussed in section \ref{subsec:fit}). 

\subsection{Manual adjustments}
\label{subsec:man}
The process of creating reference points in order to fit the model partially relies on the extraction of traces and spectra using the model itself. It is therefore important that the initial model we start with is already relatively close to the final model, in order to avoid mis-identification of shutters or lines if the offset is too large. Therefore, we have developed routines in the form of Python GUIs that allow a rough adjustment of the model parameters in a manual procedure, without having to extract every source and perform the (time-consuming) full fit. 

Manual adjustments of the model are done for imaging mode by adjusting the mirror angles dependent on the positions of the slit images on the detector, and for the spectroscopic mode in dispersion and cross-dispersion direction by manually moving extraction windows on-top of traces and line positions in spectra.

\subsection{Extraction of the reference points}
\label{subsec:ext}

The next step in the fit of the spectrograph model is the extraction of the reference points (i.e. the measured points on the detector). We have developed and optimized methods to obtain highly accurate centroid positions on the detector, which are different for imaging or spectroscopic mode data. 

\subsubsection{Imaging reference points}

For the imaging reference points, we use a 3x3 checkerboard MSA configuration. This provides a good compromise between having as many points as possible with the best spatial coverage, and avoiding crowding of the shutters that would complicate the centroid measurements. We use Source Extractor [\citenum{Bertin96}] to find and extract the centroids of the shutter images on the detector. The extraction parameters were optimized for the microshutters' shape and contrast. We then cross-reference the shutter centroids with a shutter operability file to avoid having contaminated shutters and mis-identifications in the list. This results in a cleaner cut of the extracted shutter positions. 

\subsubsection{Spectral reference points}

\begin{figure} [ht]
	\begin{center}
		\begin{tabular}{c} 
			\includegraphics[width=0.5\textwidth]{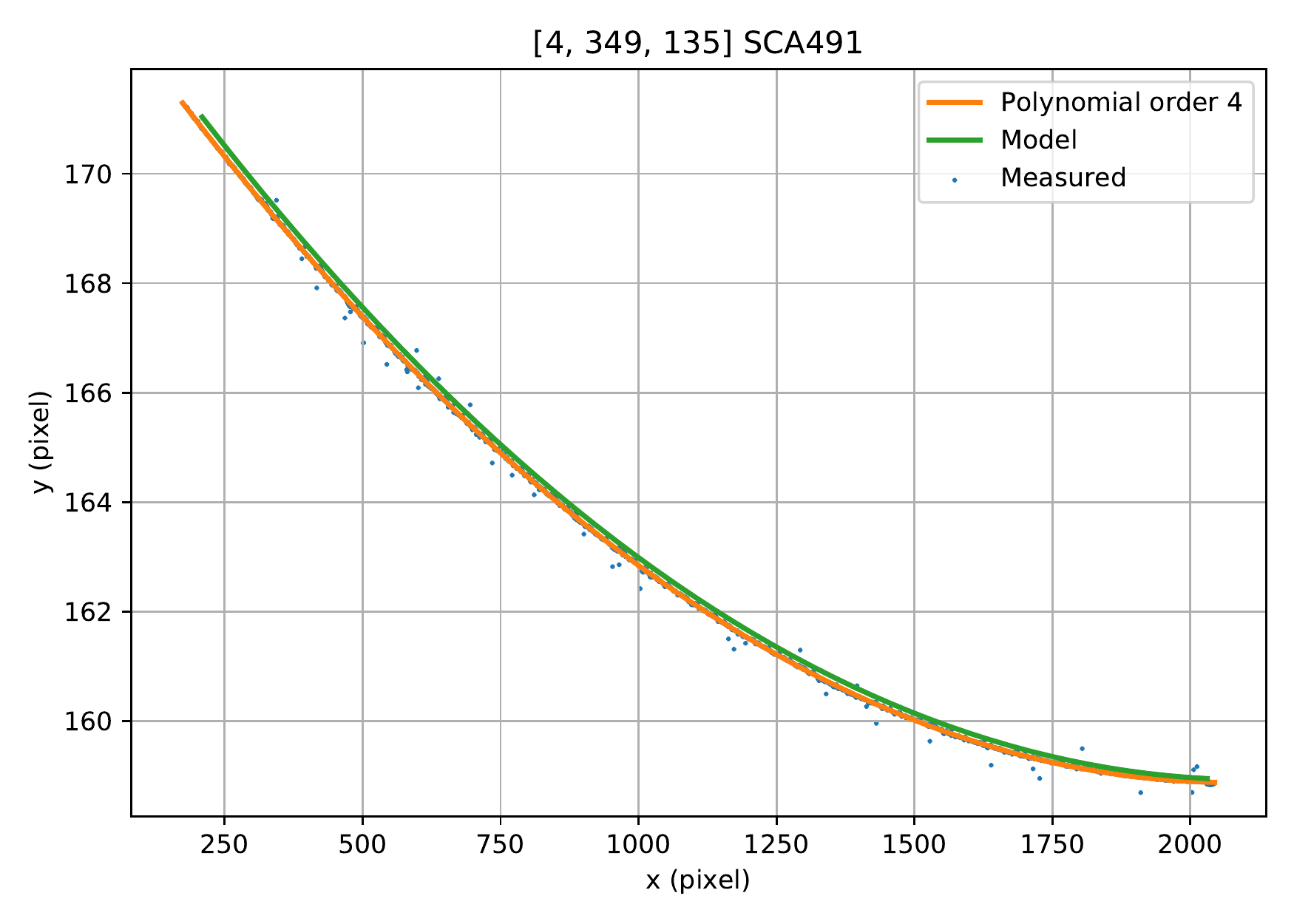}
		\end{tabular}
	\end{center}
	\caption{Example for a polynomial fit to a spectral trace on the detector, in this case for shutter (349,135) in MSA quadrant 4. The y-values were derived by computing the weighted mean for each pixel column on the detector. The polynomial fit to the data (orange line) is compared to the trace position predicted by the current model (green line). } 
	\label{fig:polyfit} 
\end{figure}

For the spectral reference points, two steps are needed: 1) tracing the spectra on the detector, and 2) measuring the center of emission lines in the spectral direction. To trace the spectra on the detector, we use exposures taken with the FLAT calibration lamps. Those calibration lamps include five continuum sources intended for spectral flat-fielding that employ filters matching the five long-pass order-separation filters in the FWA (see [\citenum{Jakobsen22}]). The spectra on the detector are captured within a so-called `extraction window', and the flux-weighted centroid of light is calculated along each detector column (the y-direction), thus defining the `trace' as a function of column number (the x-direction). A 4th-degree polynomial (3rd-degree for the prism due to its shorter traces) is fit to the trace for each aperture and exposure (see figure \ref{fig:polyfit}). These polynomials are later used to obtain the exact (x,y) position of a given reference point on the detector.

\begin{figure} [ht]
	\begin{center}
		\begin{tabular}{c} 
			\includegraphics[width=0.45\textwidth]{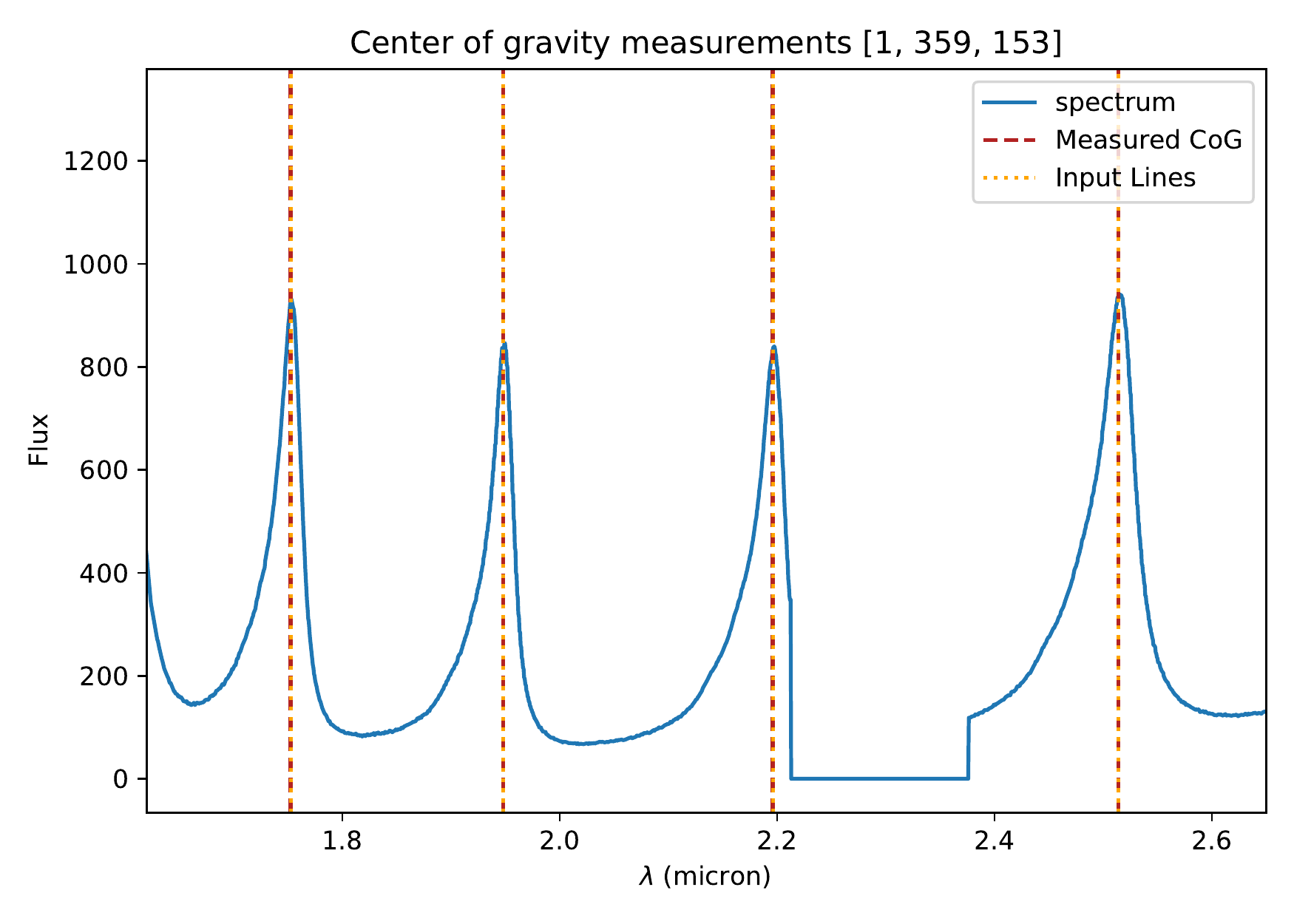}
			\includegraphics[width=0.45\textwidth]{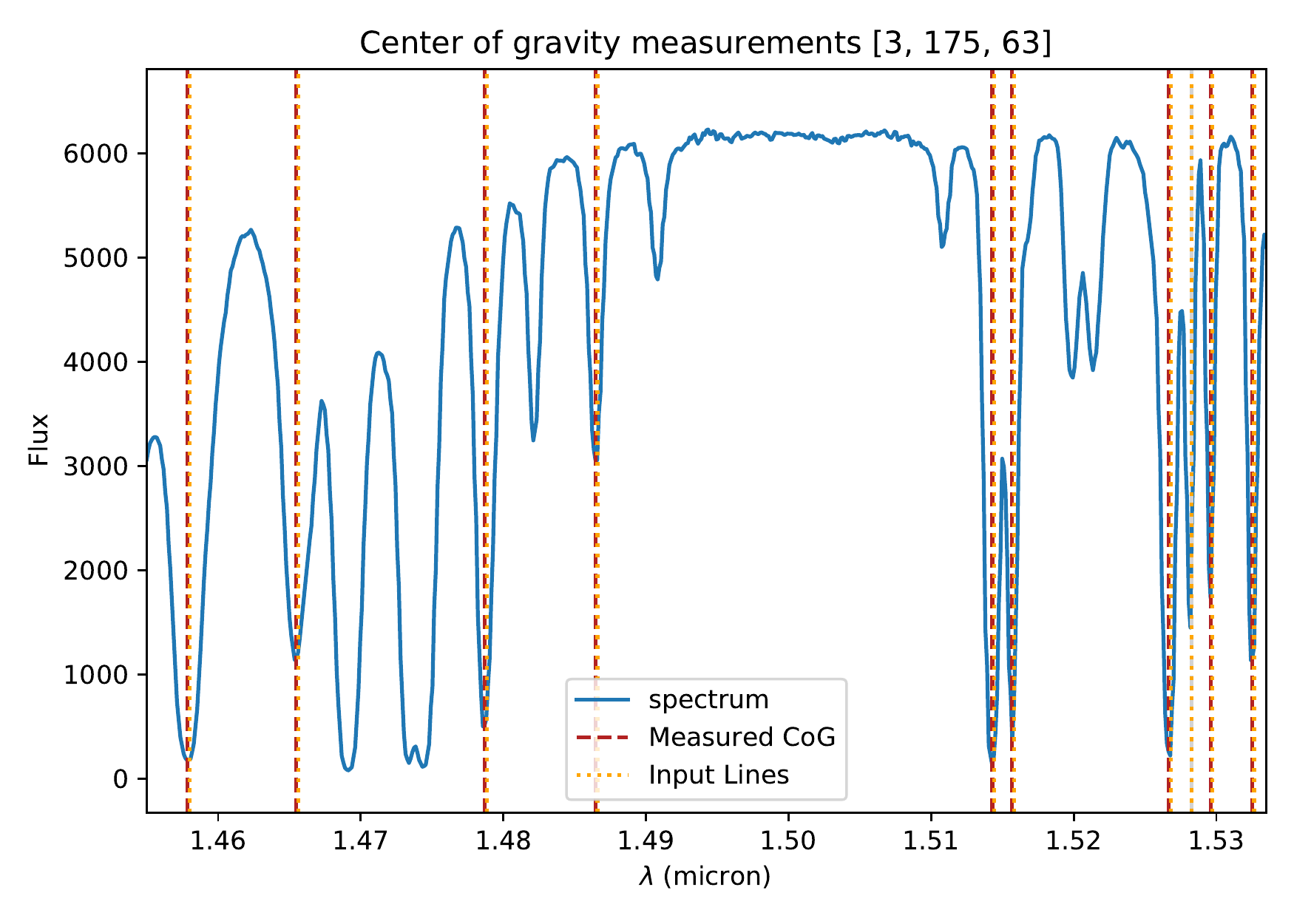}
		\end{tabular}
	\end{center}
	\caption{LINE and REF lamp spectra used for the extraction of reference points in the model fit. The derived centers of gravity are depicted by red dashed lines while the predicted line positions are shown with the yellow dotted lines.} 
	\label{fig:lines} 
\end{figure}

For the spectral reference points, we use the LINE calibration lamps, most of which are Fabry-Perot type interference filters that produce 5 to 6 Lorentzian-profiled wavelength calibration lines in each of NIRSpec bands, thus covering the full $0.6-5.3 \mu $m spectral range. A separate calibration lamp (REF) uses an Erbium-doped filter that provides an absolute wavelength reference near $1.4 \mu m$ with sharp absorption features. The spectra are extracted for all modes and apertures and flat-fielded using the before-mentioned FLAT lamp exposures. Afterwards, they are rectified and extracted to a 1D spectrum using the  NIRSpec Instrument Pipeline Software (NIPS [\citenum{Alves18}]). For each spectrum, the centroid positions of the emission and absorption lines are extracted using the center of gravity approach [\citenum{Cameron89}] with a peak fraction of ft=0.6 and customized extraction windows for each line, based on previous measurements and knowledge of the line positions (see Fig.\,\ref{fig:lines}).

The final set of spectral reference points is computed by determining the position of all spectral line on the detector, using the measured wavelengths and the polynomials derived from the traces. Note that this approach uses the model under the assumption that the spatial variations of the distortions are small and smooth. The process is repeated after the first model fit, to allow for an iterative approach of the reference point creation. 

\subsection{The imaging model}

For very early commissioning activities such as creation of the MSA operability files (see [\citenum{Rawle22}]), a model for the imaging mode was needed in order to be able to identify individual shutters. This could be achieved using TEST lamp exposures through a (3x3) checkerboard configuration of the MSA.

In this procedure, only the two mirror angles in the x and y directions (which in the full model fit are set to zero) are fit, in order to obtain the best match (least squares solution) between the positions measured on the detector and those predicted by the model. Since the final residuals were still too large, and a clear structure was being detected, we also fit the geometrical parameters of the MSA (i.e. the position and rotation of each quadrant) since they could have moved during launch. After computing the grating wheel sensor calibration for the mirror (see Section \ref{sec:gwa} and [\citenum{Alves22}]), the imaging model was complete. 

\subsection{The full spectrograph model fit}
\label{subsec:fit}

The first fit, that also contains the largest number of parameters fit at once, is performed on all the data (MOS, FS) except the dispersers G140H, G235H, and PRISM because of their rather uneven distribution of reference points on the detector. For the remaining dispersers, the grating angles $\theta_x, \theta_y$ and $\theta_z$ are fit with a few exceptions listed below. Furthermore, the MSA geometrical parameters, the FPA geometrical parameters, as well as the camera (CAM) and collimator (COL) forward transforms are fit. In total, this amounts to 118 free parameters that were fit to approximately 27600 reference points. In order to not bias the fit towards imaging data, only 1500 randomly selected microshutters per quadrant are used during the fit. For fitting, we use a Trust Region Reflective algorithm (trf) as provided in the \verb|scipy| routine \verb|least_squares|. As described in [\citenum{Dorner16}], not all parameters of the model are fit freely. There are some assumptions on symmetry and conventions that are applied and listed below:

\begin{itemize}
    \item The GWA MIRROR has all alignment tilt angles at 0, and defines the GWA reference plane. All other dispersers have alignment tilt angles relative to this surface.
    \item The MSA quadrants are rectangular and regular in size, i.e. in an individual quadrant the shutter pitch is uniform, and the shutter axes are perpendicular.
    \item The physical gap between NIRSpec's two Sensor Chip Assemblies (SCAs) is forced to be centered on the y-axis. SCA 491 has no rotation and is symmetrical to the x-axis. SCA 492 is free to move and rotate within the first condition that couples the positions of 491 and 492. This does not restrict the modeling of the instrument, as the movements can be compensated by the distortion polynomials of the CAM. It does, however, produce a geometrically simple FPA description without excessive tilts and offsets.
    \item The COL, CAM and FORE transforms are assumed to be achromatic (all-reflective optical parts) with no wavelength dependence of the distortion coefficients.
    \item The alignment tilt angle in z direction for the grating G395H ($\theta_z$) is fixed to +30 degrees (factory setting). 
\end{itemize}

The last point was an improvement to the process introduced in commissioning in order to avoid unphysical values of the grating wheel rotation angles introduced during the fit, which were then compensated by rotations in the COL and CAM transforms. By fixing the $\theta_z$ angle for one of the gratings, we allow for variations in the tilt angles of the remaining gratings relative to each other, while keeping them in a realistic range. Otherwise, the fit tends to converge on nonphysical values, due to the degeneracy with the rotations in the coordinate transforms.

\begin{figure} [ht]
	\begin{center}
		\begin{tabular}{c} 
			\includegraphics[width=\textwidth]{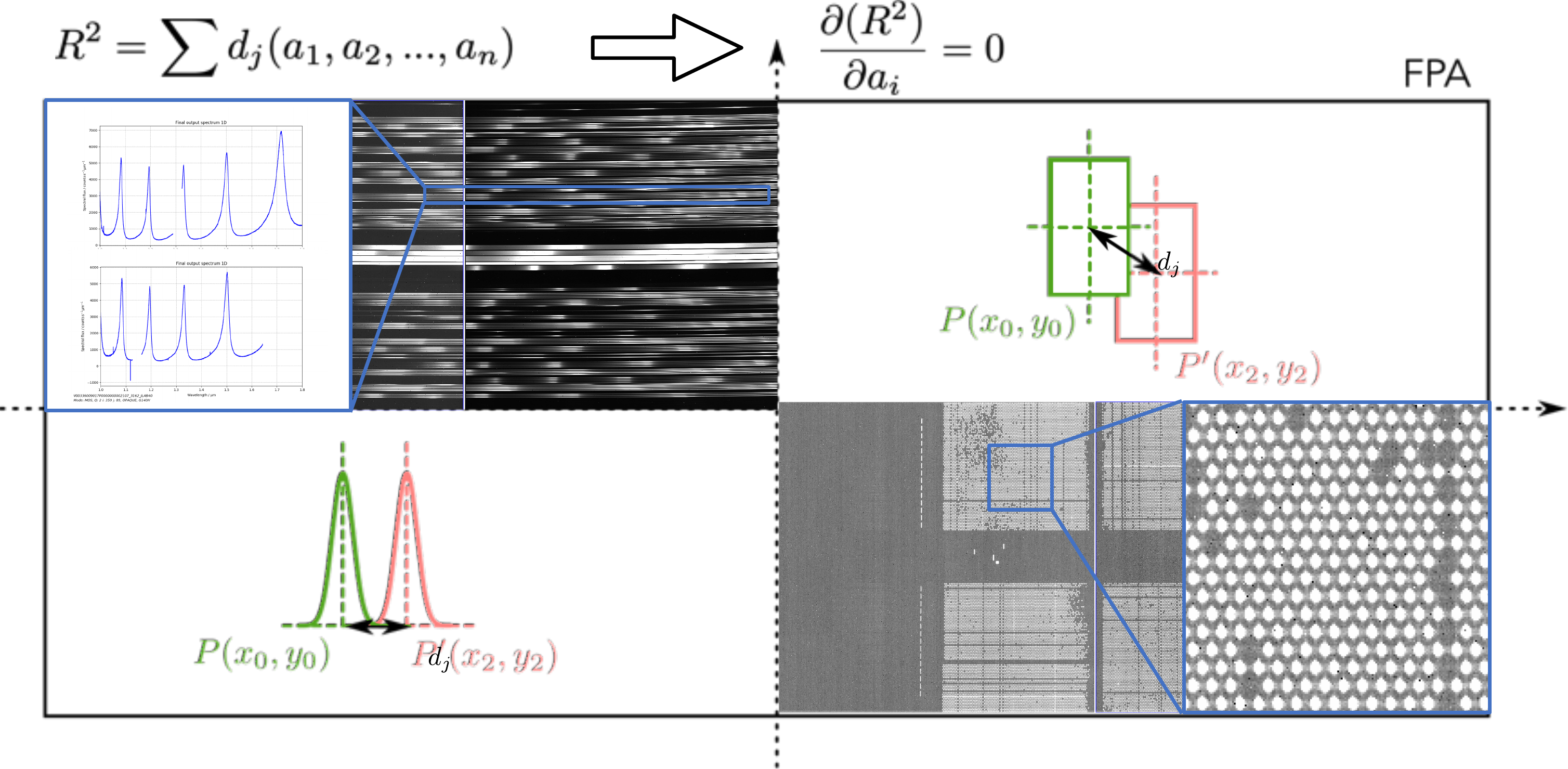}
		\end{tabular}
	\end{center}
	\caption{Concept of fitting the spectrograph model that also applied to other parts of the model fit. Reference points are extracted and compared with the predicted positions on the detector by the current model. A least square fit is applied to derive the optimal model parameters.} 
	\label{fig:spec_concept} 
\end{figure}

In a second step, the positions of the fixed slits were optimized individually, because their initial positions may not be optimal in combination with the new collimator distortion, which was dominated by the MOS data in the previous optimization ($\sim 20 \times$ more shutter references than for fixed slits). Therefore, another fitting run was done, only changing the SLIT positions in the MSA, using the reference points of the fixed slits only. The third step to complete the forward model fit, was to fit the angles of the remaining dispersers (G140H, G235H, and PRISM) individually using the newly derived distortions and slit positions from the previous fit. 

After completing the forward fit for the spectrograph model, we have now a good description of the light passing from the MSA to the detector for each mode. However, the backwards transform (going from detector to the MSA plane) cannot be analytically derived from the 5th degree polynomial transform. Therefore, a numerical approach is used to compute the backwards transforms. Here, only the collimator and camera coordinate transforms need to be derived, since all geometrical parameters remain the same in both directions. The fit does not require any exposure data, it merely uses the forward transforms for a regular grid of points. The concept is the same for either transform: for the collimator transform, a regular grid is created in the MSA plane, and the forward transform of the model is used to translate this grid to the GWA-in plane. Similarly, the camera transform is derived from a regular grid of points in the GWA-out plane, and the forward transform of the model is again used to translate this grid to the FPA plane.
In either case, a least square fit is applied to the equation $X_{out} = A X_{in}$ to obtain the polynomial coefficients of matrix $A$. 

Now that the forward and backward transforms for MOS and FS mode are in hand, the final step is to fit the positions of the virtual IFU slits on the detector. This is done by using a REF lamp exposure of the G395H grating (due to it's even distribution of the absorption lines on the detector) in combination with an IFU mirror exposure and fitting the $x_{out}, y_{out}$ positions from the IFU-POST paraxial transform for each slice. 

In the final verification step, all reference points are re-generated (re-extracted) using the new model and the residuals (difference between measured and predicted points on the detector) are computed.

\begin{figure} [ht]
	\begin{center}
		\begin{tabular}{c} 
			\includegraphics[width=\textwidth]{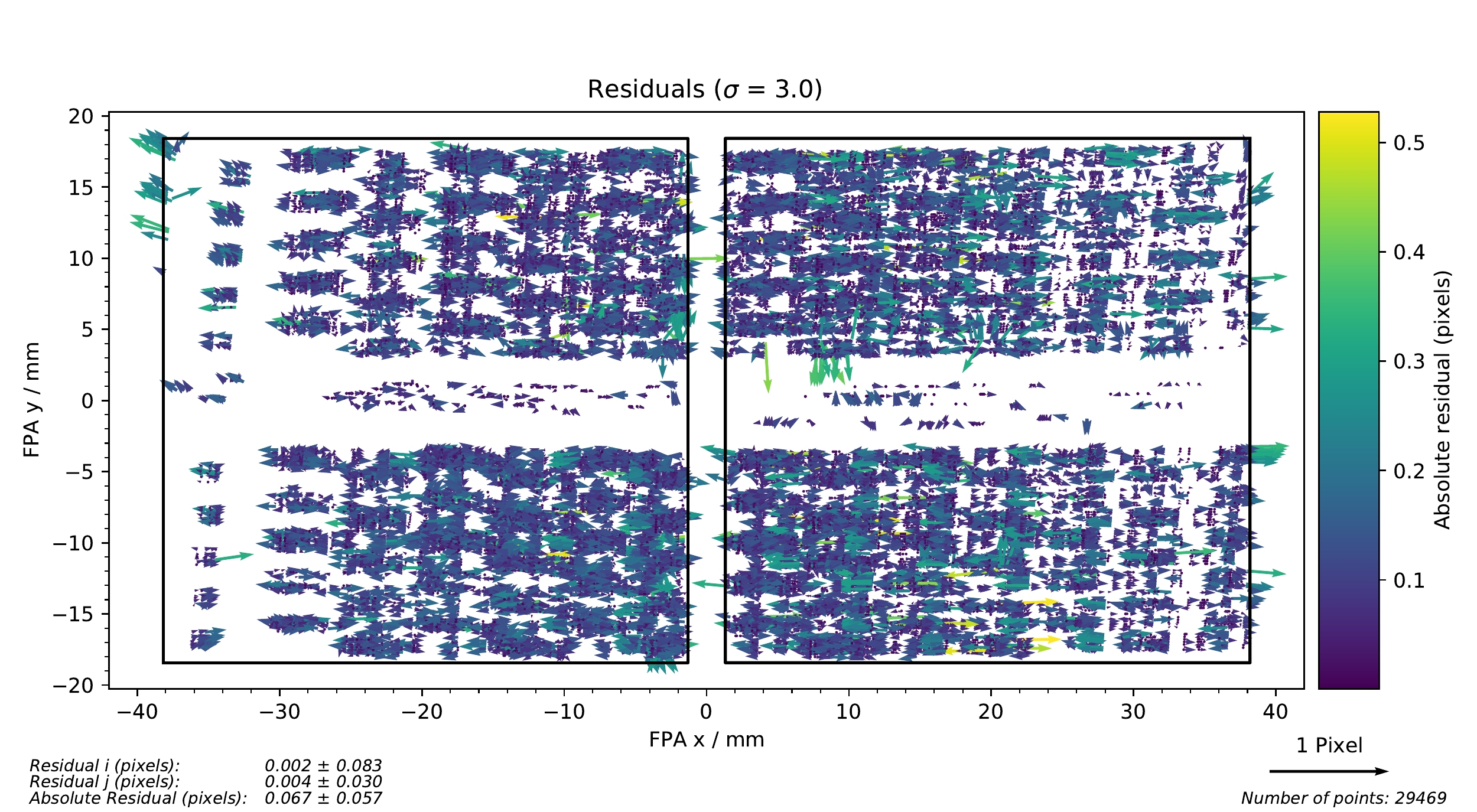}
		\end{tabular}
	\end{center}
	\caption{Residuals of forward coordinate transforms from MSA to FPA, for all gratings. The vectors indicate amplitude and direction of the difference between measured and model-computed location of the data points.} 
	\label{fig:results_spec} 
\end{figure}

Figure \ref{fig:results_spec} and Table \ref{tab:results_spec} show the results for the spectrograph model fit. All residuals remain within 0.1 pixel RMS, which should be compared to the maximum acceptable standard deviation (derived from the NIRSpec calibration error budget) of 1/10 of a resolution element or 0.2 pixels. The residuals for all modes meet this requirement with substantial margin (a factor of two). They are also very similar to the residuals derived from ground campaign data, which implies that no major components of the NIRSpec optical train have moved noticeably from their pre-launch positions.\footnote{In fact, the residuals for the prism improved by a factor of two, due to the use of a more accurate prescription for the index of refraction of $\rm CaF_2$ as a function of wavelength, compared to what was used for pre-launch model fits.}

\begin{table}[ht]
\caption{Residuals of the optimized model forward projection from MSA to FPA per grating, for MOS and FS mode. Here, $i$ is the coordinate in dispersion direction, and $j$ in spatial direction.} 
\label{tab:results_spec}
\begin{center}       
\begin{tabular}{l|r|r|r} 
\hline
\rule[-1ex]{0pt}{3.5ex}  Grating & i mean + RMS & j mean + RMS & median absolute  \\
\hline
\rule[-1ex]{0pt}{3.5ex}  G140H &  $0.011\pm0.063$ & $0.000\pm0.022$& $0.054\pm0.041$    \\
\rule[-1ex]{0pt}{3.5ex}  G140M &  $0.001\pm0.096$ & $0.003\pm0.033$& $0.066\pm0.077$   \\
\rule[-1ex]{0pt}{3.5ex}  G235H &  $0.002\pm0.102$ & $0.011\pm0.054$& $0.096\pm0.066$    \\
\rule[-1ex]{0pt}{3.5ex}  G235M &  $0.000\pm0.062$ & $0.003\pm0.021$& $0.054\pm0.036$    \\
\rule[-1ex]{0pt}{3.5ex}  G395H &  $0.001\pm0.089$ & $0.003\pm0.022$& $0.071\pm0.057$   \\
\rule[-1ex]{0pt}{3.5ex}  G395M &  $-0.001\pm0.072$ & $0.003\pm0.018$& $0.061\pm0.042$   \\
\rule[-1ex]{0pt}{3.5ex}  PRISM &  $-0.004\pm0.087$ & $0.000\pm0.022$& $0.076\pm0.048$   \\
\rule[-1ex]{0pt}{3.5ex}  MIRROR &  $0.001\pm0.012$ & $0.003\pm0.015$& $0.017\pm0.009$   \\
\hline 
\end{tabular}
\end{center}
\end{table}

\section{THE GRATING WHEEL SENSOR CALIBRATION}
\label{sec:gwa}

A detailed description of the grating wheel sensor calibration is provided in [\citenum{Alves22}]. Here, we only provide a short summary.  

The finite angular positioning repeatability of the grating wheel causes small but measurable displacements of the light beam in the focal plane, thus ruling out a static solution to predict the light-path. To address that, the GWA includes two magneto-resistive position sensors that are read after every wheel move, in order to measure the precise ‘tip and tilt’ orientation of each GWA element. 

However, the precise relation between the measured sensor voltage and actual wheel position must be carefully calibrated after every cryo-cycle of NIRSpec. In practice, this is done by fitting the relationship between the sensor voltages and the observed angular displacement of the wheel, measured relative to the selected reference point of the spectrograph instrument model. For the dispersers, a number of internal lamp exposures with intermediate movements of the grating wheel are taken. Absorption lines in the spectra, as well as shutter centroids in an image of the (3x3) checkerboard MSA configuration are used to measure the absolute positions, and to derive the actual grating and mirror angles as described in sections \ref{subsec:ext} and \ref{subsec:fit}. The calibration relations between the sensor voltage reading and the measured angular displacement have been shown to be linear, both for the MIRROR and all dispersers. For every NIRSpec exposure, they are then applied to complete the geometrical description of the state of the instrument.

\section{ASTROMETRIC CALIBRATION}
\label{sec:fore}

The final step to complete the full NIRSpec instrument model is the astrometric calibration or FORE optics fit. The FORE transform can be measured from images of the sky, but it should be kept in mind that the optical path from the sky to the MSA plane includes the OTE optics (see Fig,\,\ref{fig:scheme}, and discussion below). Because the refractive properties of the various filters in the FWA are different, the transform needs to be fit separately for each filter. In order to obtain these fits, two imaging mode exposures of the JWST astrometric reference field in the Large Magellanic Cloud (LMC) through the ALLOPEN shutter configuration were taken for each filter.  

\begin{figure} [ht]
	\begin{center}
		\begin{tabular}{c} 
			\includegraphics[width=\textwidth]{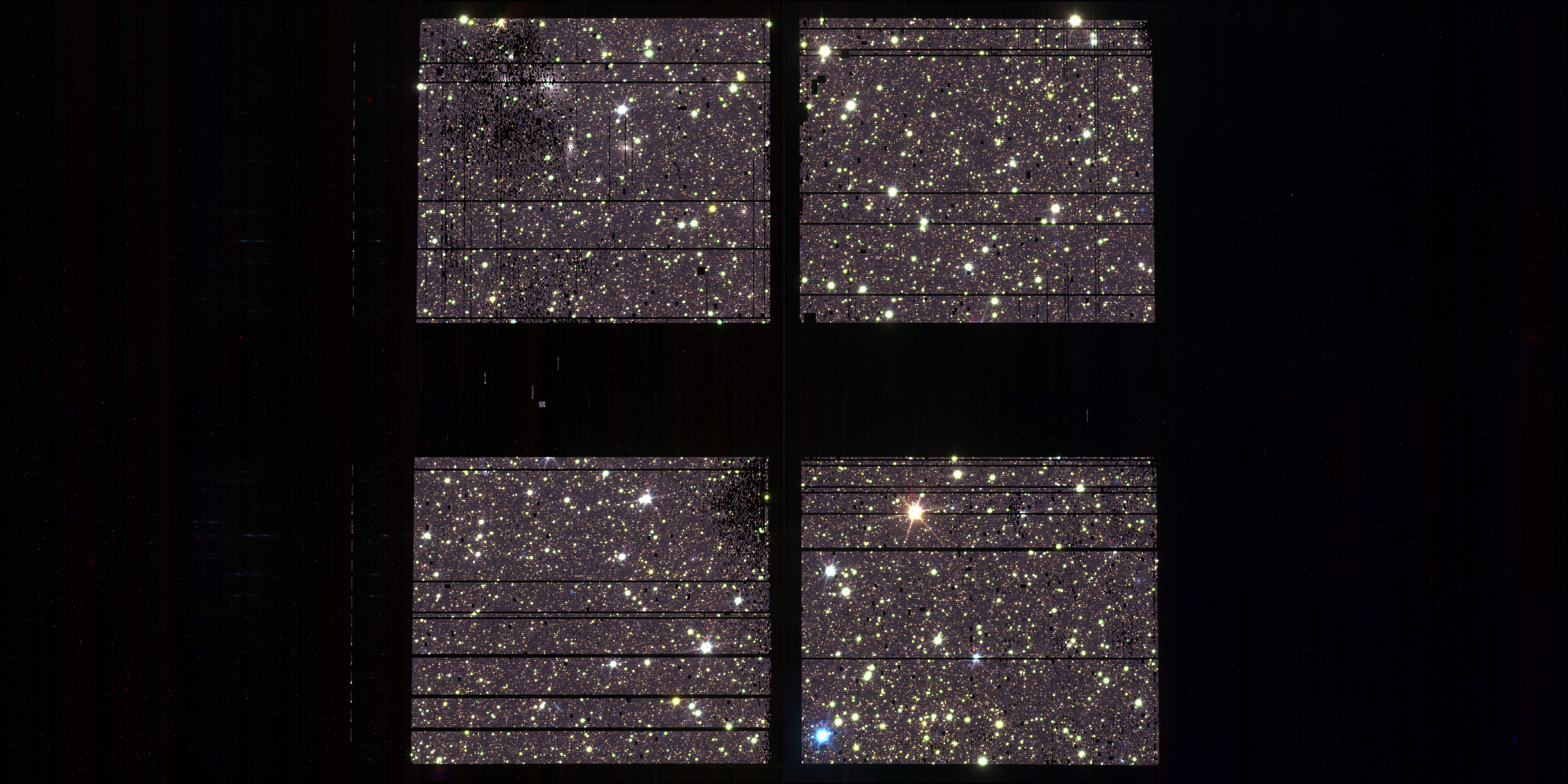}
		\end{tabular}
	\end{center}
	\caption{Color image of the LMC astrometric field viewed through the open shutters on the NIRSpec detector.} 
	\label{fig:astro} 
\end{figure}

Figure \ref{fig:astro} shows an example exposure of the astrometric field viewed through the open shutters. It is obviously a very crowded field which makes identification of individual stars challenging. The problem is made worse by the fact that the grid of shutter bars and the various failed-closed shutters affect the shape of the point spread function (PSF) of individual stars. In order to mitigate these effects, the two exposures per filter were `dithered', i.e. offset by half the shutter pitch in both x- and y-direction.

\begin{figure} [ht]
	\begin{center}
		\begin{tabular}{c} 
			\includegraphics[width=\textwidth]{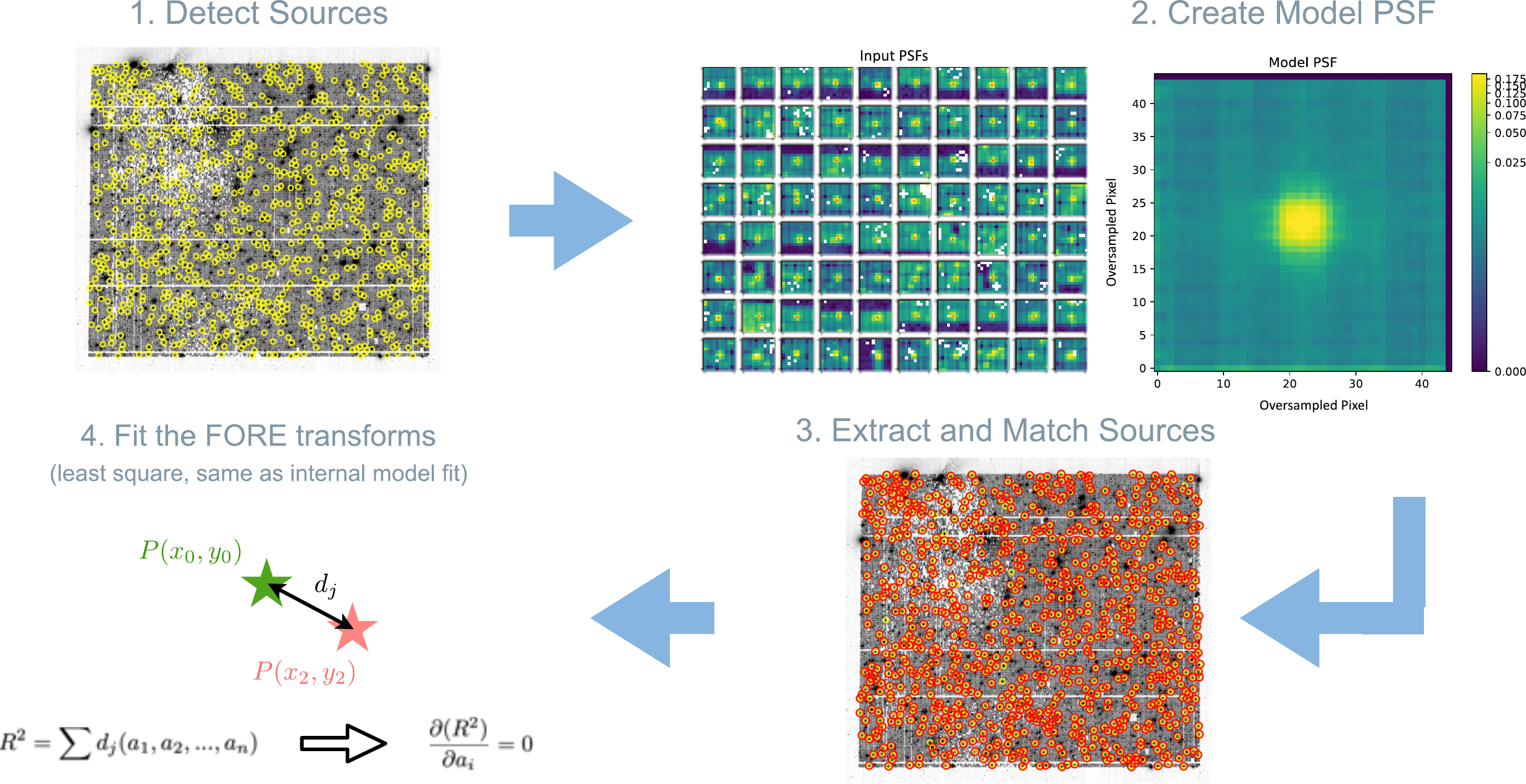}
		\end{tabular}
	\end{center}
	\caption{Schematic outline of the astrometric calibration. In short: positions of stars are measured on the detector and matched with the astrometric catalogue. The instrument model is used to predict the star positions on the detector from their sky coordinates in the stellar catalogue. A least-squares fit of the predictions to the measurements is then applied to obtain the optimal model parameters.} 
	\label{fig:fore_concept} 
\end{figure}

We used a python implementation of \verb|DAOFIND| [\citenum{Stetson87}] with parameters optimized to the NIRSpec data, to detect stars and record their positions. A subset of those stars meet certain isolation and roundness criteria, and are used to obtain a model PSF. This is done in an iterative algorithm as provided in the \verb|photutils| python package. An example of a final model PSF is shown in figure \ref{fig:fore_concept}. Using the PSF model, the centroids of all stars are extracted over the full field of view, for each filter and for both dither positions. 

The fit of the FORE transform for each filter is, in essence, a fit of the combined (OTE + FORE) transform, but since the OTE transform is fixed, only the FORE parameters are fit. This was done as follows: first, the paraxial parameters of the transform (the reference points in the input and output plane, the magnification factors and the rotation) are set to unity and only the transforms are fit. The new paraxial parameters are extracted using the transform itself and its derivatives. The distortions are then fit again while fixing the new paraxial parameters. In this way, we ensure that only meaningful paraxial parameters are used, and only the effects of distortion are captured in the transforms. On average, each filter had about 5000 stars used for the fit. The JWST Telescope’s team used observations from FGS1 and FGS2 obtained in parallel mode, to determine the most accurate pointing and roll angle information for each of the NIRSpec calibration exposures. This was important, as the uncertainty in pointing and roll would have been absorbed in our model fit, thus resulting in an incorrect model for any other pointing on the sky.

Note that we perform the fit of the FORE transforms using coordinates measured on the detector, even though the output plane of the FORE transform is the MSA. This requires fixing all the other transforms (OTE, COL and CAM) by using the newly fit internal model from section \ref{sec:spec}. We do not use the grating wheel calibration in this fit, but rather take advantage of the extra internal exposures that were taken during the astrometric calibration (see section \ref{subsec:commdata}) to derive the exact grating wheel angle for the sky exposures. This removes any uncertainties in the sensor wheel calibration. The backwards fit was performed in the same method as described for the collimator and camera transforms in Section\,\ref{sec:spec}. 

\begin{figure} [ht]
	\begin{center}
		\begin{tabular}{c} 
			\includegraphics[width=\textwidth]{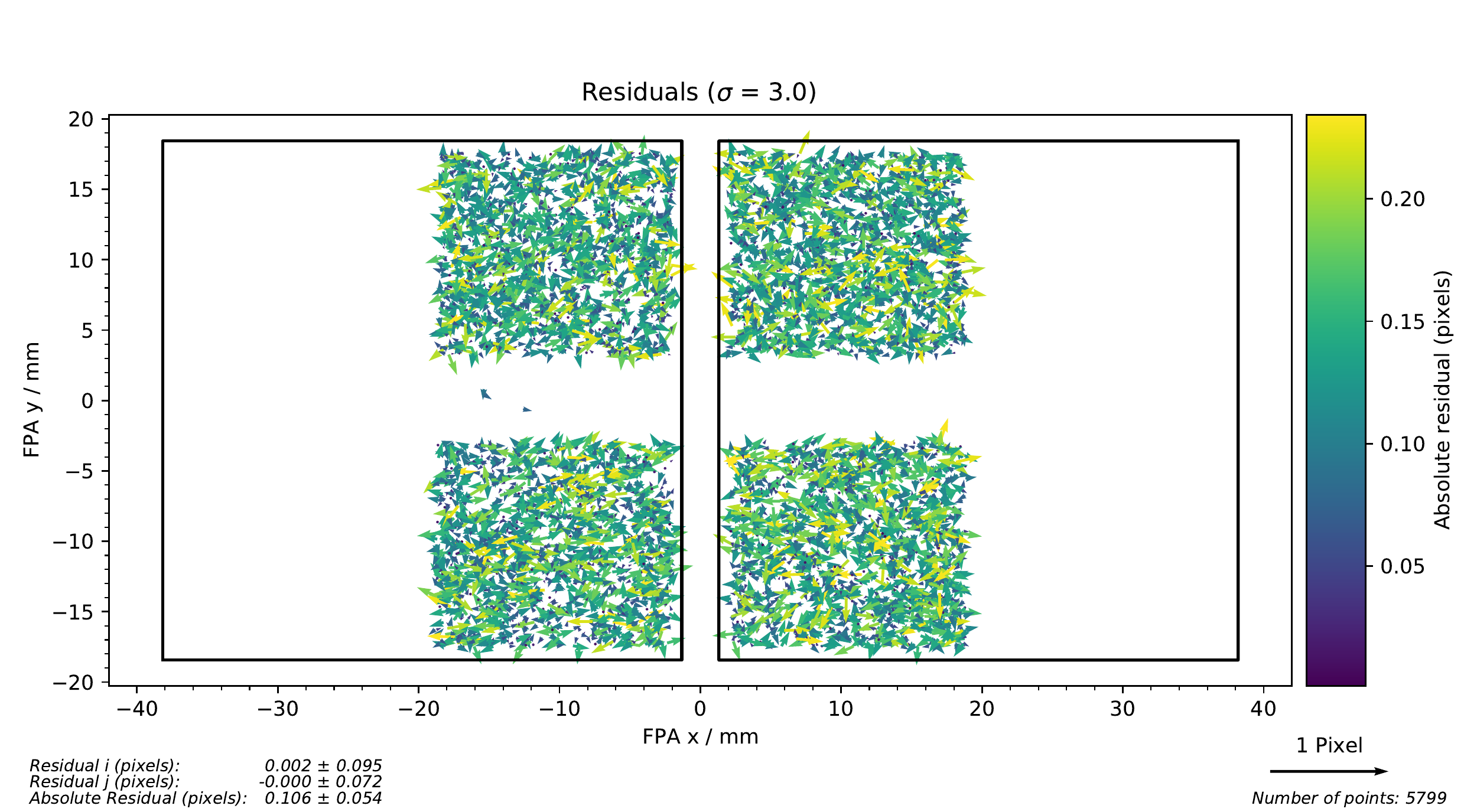}
		\end{tabular}
	\end{center}
	\caption{Residuals of the FORE forward coordinate transforms from sky to MSA, for the F110W filter on the detector. The vectors give amplitude and direction of the difference between measured and model-computed location of the data points.} 
	\label{fig:fore_results} 
\end{figure}

Figure \ref{fig:fore_results} shows the residuals for one filter (F110W). Table \ref{tab:fore_results} shows the results for all filters.  Residuals for all filters are range between 0.1-0.25 pixel RMS. This is slightly larger than what we expected from using simulations ($<0.1$). As mentioned before, the bars of the shutters and the crowding in the field make the centroiding process challenging. We therefore assume that the larger uncertainties mainly originate from the uncertainty in the centroiding, rather than from the accuracy of the model, and performed several tests to confirm this, described in the following section.

\begin{table}[ht]
\caption{Residuals on the detectors of the optimized model forward projection for the FORE transform per filter. Here i is for dispersion direction and j for spatial direction.} 
\label{tab:fore_results}
\begin{center}       
\begin{tabular}{l|r|r|r} 
\hline
\rule[-1ex]{0pt}{3.5ex}  Filter & i mean + RMS & j mean + RMS & median absolute  \\
\hline
\rule[-1ex]{0pt}{3.5ex}  CLEAR &  $0.002\pm0.142$ & $0.000\pm0.206$& $0.156\pm0.084$    \\
\rule[-1ex]{0pt}{3.5ex}  F070LP &  $0.002\pm0.145$ & $0.000\pm0.115$& $0.163\pm0.087$   \\
\rule[-1ex]{0pt}{3.5ex}  F100LP &  $0.003\pm0.155$ & $0.000\pm0.119$& $0.174\pm0.090$    \\
\rule[-1ex]{0pt}{3.5ex}  F110W &  $0.002\pm0.095$ & $0.000\pm0.072$& $0.106\pm0.054$    \\
\rule[-1ex]{0pt}{3.5ex}  F140X &  $0.002\pm0.116$ & $0.000\pm0.086$& $0.129\pm0.066$   \\
\rule[-1ex]{0pt}{3.5ex}  F170LP &  $0.003\pm0.161$ & $0.000\pm0.122$& $0.179\pm0.095$   \\
\rule[-1ex]{0pt}{3.5ex}  F290LP &  $0.000\pm0.236$ & $-0.001\pm0.160$& $0.250\pm0.136$   \\
\hline 
\end{tabular}
\end{center}
\end{table}

\subsection{Tests on centroiding accuracy}

\begin{figure} [ht]
	\begin{center}
		\begin{tabular}{c} 
			\includegraphics[width=\textwidth]{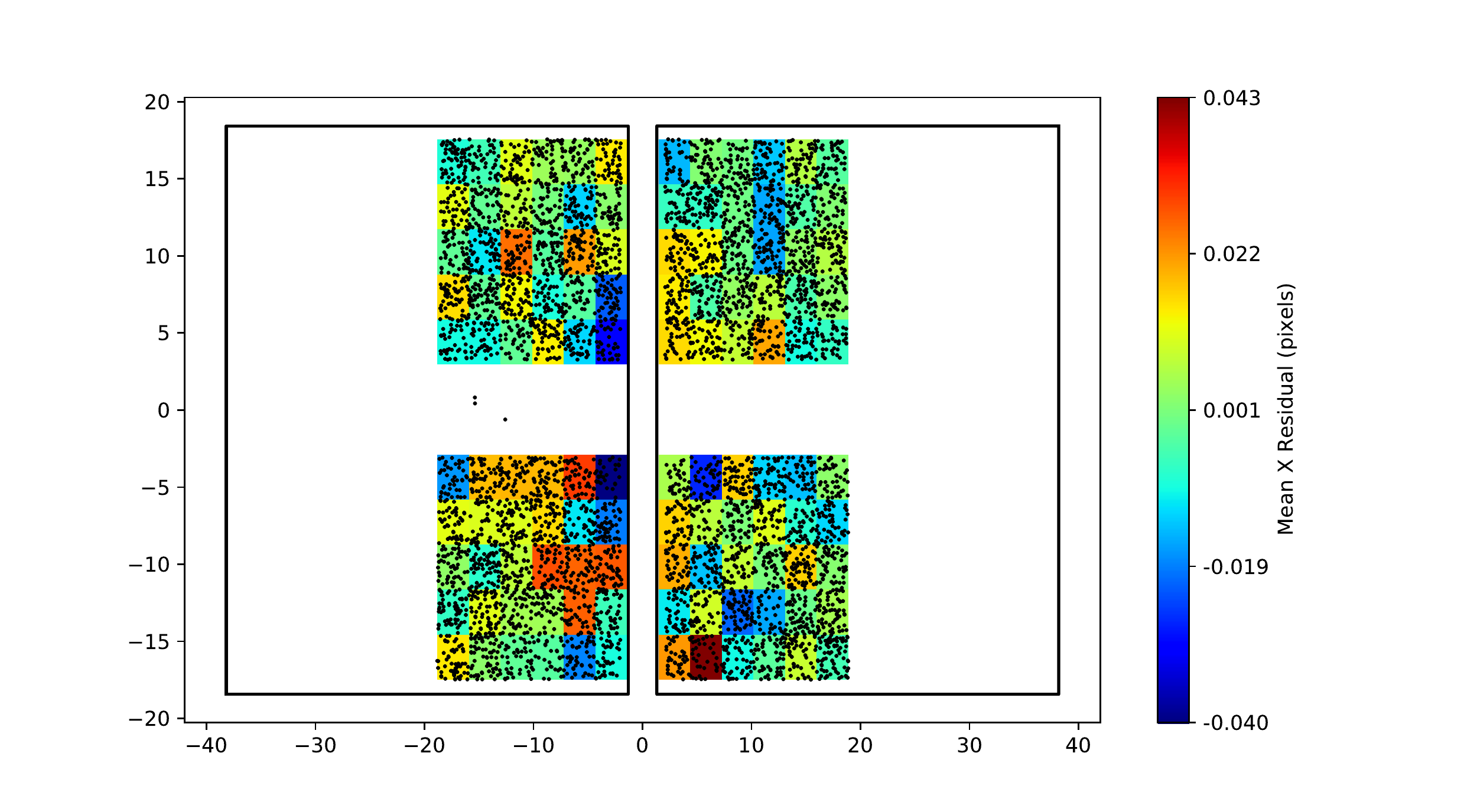}
		\end{tabular}
	\end{center}
\caption{Binning the spatial residuals of the F110W filter over the field of view in order to test for gradients or spatial dependencies. The distributions of the individual bins is symmetric around zero} 
\label{fig:sections} 
\end{figure}

In order to test the accuracy of our centroiding algorithm and the smoothness of the residuals we performed two tests. The first one was to divide the residuals into different sections on the detector, and to measure their mean and RMS within each section. This allows us to verify that there are no strong spatial variations in the residuals, which would hint towards residual distortions in the model. Figure \ref{fig:sections} shows one of those tests: the distribution appears to be symmetric around zero, smooth over the field of view and without any spatially organised structure present.

\begin{figure} [ht]
	\begin{center}
		\begin{tabular}{c} 
			\includegraphics[width=\textwidth]{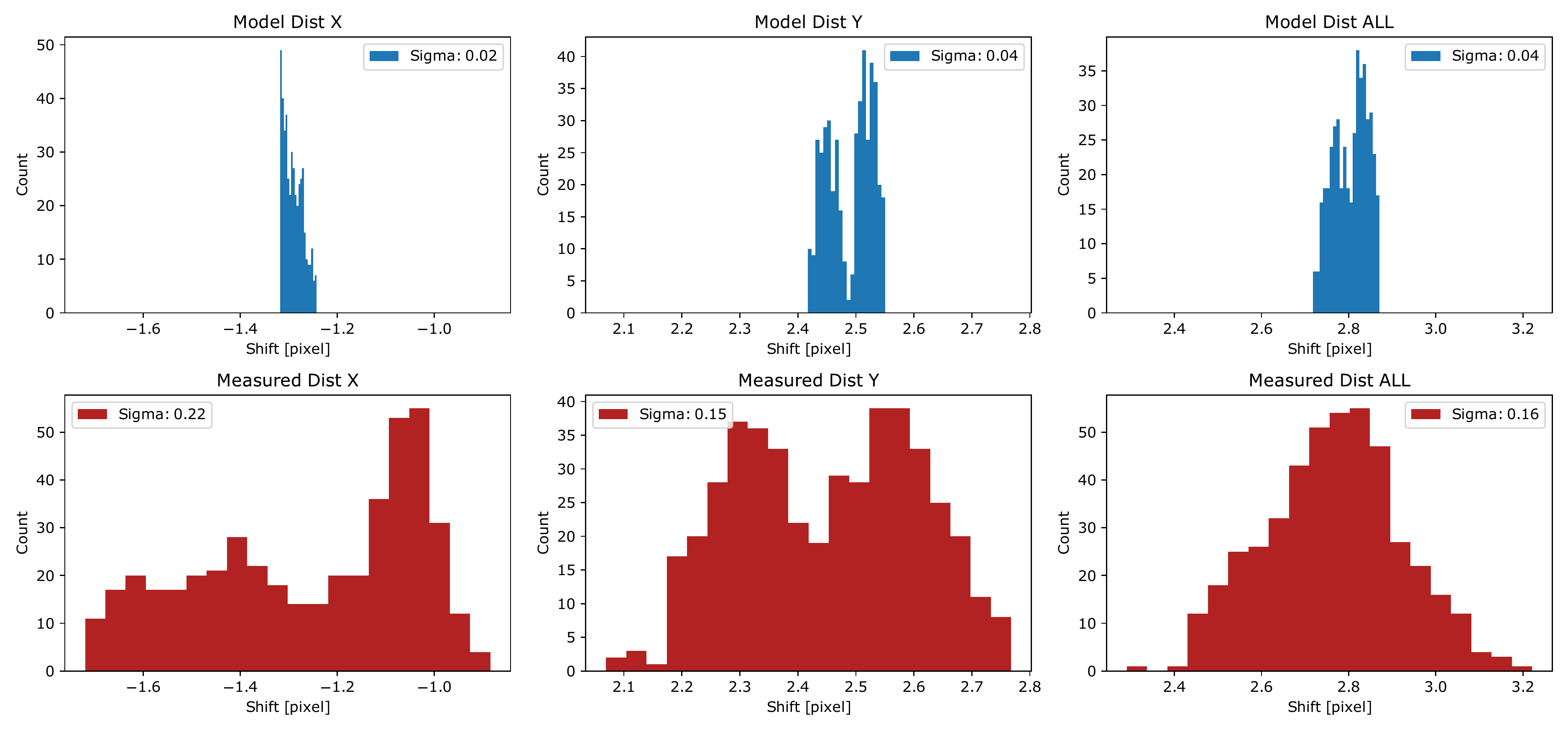}
		\end{tabular}
	\end{center}
\caption{Test of centroiding accuracy, performed by locating the same stars in two dithered images. The upper panels show the distribution of shifts as calculated by the model. The width and shape of the distribution is due to the distortions in the instrument and the spatial distribution of the four quadrants. The lower panels show the measured shifts on the detector. It is clear that the wider shape of those distributions compared to the modelled ones is caused by the limited centroiding accuracy. } 
\label{fig:dithers} 
\end{figure}

The second test made use of the fact that we have two dither positions for each filter. By comparing the measured shift for a given star with the one predicted by the model, and analyzing the distribution for all stars, we can estimate the average spread in the centroiding distribution. Figure \ref{fig:dithers} shows the results of this test. It is clear that the spread in positions is of the order of 0.15 pixel in either direction, which would explain the higher spread in the residuals compared to the simulations.

Given the results from these tests, we are confident that the larger residuals in the FORE model fit are due to the limited accuracy of our centroiding algorithm, and not due to the performance of the model. In the future, it might be worth exploring more advanced centroiding algorithms to mitigate this source of uncertainty.

\section{SUMMARY/CONCLUSIONS}
\label{sec:summary}

The NIRSpec instrument model describes the light path through the instrument from sky to the detector and back. It consists of nearly 1000 free parameters and is divided into three main parts: the spectrograph/internal model, the grating wheel sensor calibration, and the astrometric or FORE calibration. 

During commissioning campaign between December 25, 2021 and July 15, 2022, we took various data sets to obtain a final fit of the instrument model, and to provide accurate wavelength and astrometric calibration of the instrument. The data contains a large set of internal calibration lamps for each mode and grating/filter combination, as well as external imaging mode sky observations of an astrometric reference field. 

The residuals for the internal model and the grating wheel sensor calibration were as expected and similar to the ones obtained during ground campaigns. They were below the requirements of 1/10 of a resolution element (1/20 of a pixel) for all modes and gratings. For the astrometric calibration, we obtained higher residuals than expected, but concluded after performing multiple tests that they stem from the systematic uncertainty of the centroiding rather than the quality of the model fit.

External tests on the wavelength calibration and successful target acquisition using the MSA have shown that the new model derived during commissioning is in excellent shape and within all requirements. 

\subsection{Maintenance and future work} \label{subsec:future}

In a stable orbit and without further vibrations and temperature changes, the model should remain stable over time. However, this has never been tested and the grating wheel sensor showed a strong dependence on temperature during ground campaigns and cool-down phase. Therefore we have planned monitoring programs in the cycle-1 calibration campaign of JWST. This program contains a subset of the internal lamp exposures in order to validate the accuracy of the model. If there is reason to assume a substantial change to the model, contingency plans need to be explored and the model likely needs to be re-fit.  

\bibliography{report} 
\bibliographystyle{spiebib} 

\end{document}